\RequirePackage{ifpdf}
\ifpdf 
\documentclass[pdftex]{sigma}
\else
\documentclass{sigma}
\fi

\begin{document}

\renewcommand{\PaperNumber}{014}

\FirstPageHeading

\renewcommand{\thefootnote}{$\star$}

\ShortArticleName{Symmetry Transformation in Extended Phase Space:
the Harmonic Oscillator}

\ArticleName{Symmetry Transformation\\ in Extended Phase Space:
the Harmonic Oscillator\\ in the Husimi
Representation\footnote{This paper is a contribution to the
Proceedings of the Seventh International Conference ``Symmetry in
Nonlinear Mathematical Physics'' (June 24--30, 2007, Kyiv,
Ukraine). The full collection is available at
\href{http://www.emis.de/journals/SIGMA/symmetry2007.html}{http://www.emis.de/journals/SIGMA/symmetry2007.html}}}

\Author{Samira BAHRAMI~$^{\dag}$ and Sadolah NASIRI~$^{\ddag}$}

\AuthorNameForHeading{S.~Bahrami and S.~Nasiri}

\Address{$^{\dag}$~Department of Physics, Zanjan University,
Zanjan, Iran}
\EmailD{\href{mailto:bahrami.samira@gmail.com}{bahrami.samira@gmail.com}}

\Address{$^{\ddag}$~Institute for Advanced Studies in Basic
Sciences, Iran}
\EmailD{\href{mailto:nasiri@iasbs.ac.ir}{nasiri@iasbs.ac.ir}}

\ArticleDates{Received October 08, 2007, in f\/inal form January
23, 2008; Published online February 04, 2008}

\Abstract{In a previous work the concept of quantum potential is
generalized into extended phase space (EPS) for a particle in
linear and harmonic potentials. It was shown there that in
contrast to the Schr\"{o}dinger quantum mechanics by an
appropriate extended canonical transformation one can obtain the
Wigner representation of phase space quantum mecha\-nics in which
the quantum potential is removed from dynamical equation. In other
words, one still has the form invariance of the ordinary
Hamilton--Jacobi equation  in this representation. The situation,
mathematically, is similar to the disappearance of the centrifugal
potential in going from the spherical to the Cartesian
coordinates. Here we show that the Husimi representation is
another possible representation where the quantum potential for
the harmonic potential disappears and the modif\/ied
Hamilton--Jacobi equation reduces to the familiar classical form.
This happens when the parameter in the Husimi transformation
assumes a specif\/ic value corresponding to $Q$-function.}

\Keywords{Hamilton--Jacobi equation; quantum potential; Husimi
function; extended phase space}

\Classification{81S30}

\section{Introduction}

 According to the Bohm approach to quantum mechanics the quantum
 potential in the modif\/ied Hamilton--Jacobi equation may be equally
 looked at from the point of view of the Newton second law as a
 quantum force term~\cite{Bohm}. Thus, in the causal interpretation, in
 addition to the external force, the quantum force derived from
 quantum potential, guides the trajectory of the quantum
 particle~\cite{Holland}. Takabayashi~\cite{Takabayashi} and Muga~\cite{Muga} introduced the concept
 of quantum internal energy (or stress) as a consequence of the
 projection from the phase space representation to the
 conf\/iguration space representation. They argued that unlike the
 classical systems which have the kinetic and potential energies,
 quantum systems also have intrinsic internal energies associated
 with spatial localization and momentum dispersion emerging
 from their inherent extended natures suggesting a link to the
 Heisenberg position-momentum uncertainty principle~\cite{Brown}. Holland~\cite{Holland1}
  investigated the de Broglie--Bohm law of motion using a
 variational formulation. He considered a quantum system and
 suggested a total Lagrangian for the interaction of a point like
 particle with the Schr\"{o}dinger f\/ield. The interaction of the
 particle and f\/ield is attributed to a scalar potential that
 turns out to be the quantum potential. The connection of the quantum
 potential with quantum f\/luctuations and quantum geometry in
 terms of Weyl's curvature has been studied by F.~Shojai and A.~Shojai~\cite{Shojai} and Carroll~\cite{Carroll}. Nevertheless, unlike the external
 potential, the quantum potential is not a pre-assigned function
 of the system coordinates and can only be derived from the wave
 function of the system~\cite{Brown} or from the corresponding quantum
 distribution functions used to calculate the average values of
 the observables~\cite{Nasiri}. This representation dependent property of the
 quantum potential allows one to f\/ind appropriate representations
 where the quantum potential could be removed from the
 modif\/ied Hamilton--Jacobi equation. Carroll~\cite{Carroll1} has shown that
 there are generalized quantum theories for which the quantum
 potential depends on the wave function. Using the extended phase
 space formulation of quantum mechanics~\cite{Sobouti,Nasiri1,Nasiri2},
 Nasiri~\cite{Nasiri} has shown that in the Wigner representation of phase
 space quantum mechanics~\cite{Wigner} the quantum potential is
 removed from the dynamical equation of a particle in linear and
 harmonic potentials keeping the Hamilton--Jacobi equation form
 invariant.  It seems that the Husimi representation could be another
 candidate to release the quantum potential. In Section~\ref{sec2}, the EPS formalism is
 reviewed. In Section~\ref{sec3}, the extended canonical transformations
 are introduced and are used to obtain the Husimi equation and
 the corresponding solution. In Section~\ref{sec4}, an expression for the
 quantum potential is obtained in the EPS. In Section~\ref{sec5}, an
 appropriate phase space representation is found in which the
 quantum potential is removed for the harmonic potential. Section~\ref{sec6}
 is devoted to the concluding remarks.

\section{Review of EPS formalism}\label{sec2}

Assuming the phase space coordinates $p$ and $q$ to be independent
variables on virtual trajectories allows one to def\/ine momenta
$\pi_{p}$ and $\pi_{q}$, conjugate to $p$ and $q$, respectively.
One may def\/ine an extended Lagrangian in the phase space as
follows~\cite{Sobouti,Nasiri1}
\begin{equation}
{\cal L}(p,q,\dot{p},\dot{q})=-\dot{p}q-\dot{q}p+{\cal
L}^{p}(p,\dot{p})+{\cal L}^{q}(q,\dot{q}),\label{eq1}
\end{equation}
where ${\cal L}^{q}$ and ${\cal L}^{p}$ are $q$ and $p$ space
Lagrangians, satisfying the following Legendre transfor\-ma\-tion,
respectively,
\begin{gather*}
H\left(\frac{\partial {\cal L}^{q}}{\partial \dot{q}},q\right)=\dot{q}\frac{\partial {\cal L}^{q}}{\partial \dot{q}}- {\cal L}^{q}(q,\dot{q}),\qquad 
H\left(p,\frac{\partial {\cal L}^{p}}{\partial
\dot{p}}\right)=-\dot{p}\frac{\partial {\cal L}^{p}}{\partial
\dot{p}}+
{\cal L}^{p}(p,\dot{p}).
\end{gather*}
The f\/irst two terms in equation~\eqref{eq1} constitute a total
time derivative. The equations of motion are
\begin{gather}
\frac{d}{dt}\frac{\partial {\cal L}}{\partial \dot{p}}-\frac{\partial {\cal L}}{\partial p}=\frac{d}{dt}\frac{\partial {\cal L}^{p}}{\partial \dot{p}}-\frac{\partial {\cal L}^{p}}{\partial p}=0,\qquad 
\frac{d}{dt}\frac{\partial {\cal L}}{\partial
\dot{q}}-\frac{\partial {\cal L}}{\partial
q}=\frac{d}{dt}\frac{\partial {\cal L}^{q}}{\partial
\dot{q}}-\frac{\partial {\cal L}^{q}}{\partial q}=0.\label{eq3} 
\end{gather}
The $p$ and $q$ in equations~\eqref{eq1} and~\eqref{eq3} are not,
in general, canonical pairs. They are so only on actual
trajectories and through a proper choice of the initial values.
This gives the freedom of introducing a second set of canonical
momenta for both~$p$ and~$q$. One does this through the extended
Lagrangian. Thus
\begin{gather*}
\pi_{p}=\frac{\partial {\cal L}}{\partial \dot{p}}=\frac{\partial {\cal L}^{p}}{\partial \dot{p}}-q,\qquad 
\pi_{q}=\frac{\partial {\cal L}}{\partial \dot{q}}=\frac{\partial
{\cal L}^{q}}{\partial \dot{q}}-p.
\end{gather*}
Evidently, $\pi_{p}$ and $\pi_{q}$ vanish on actual trajectories
and remain non zero on virtual ones. From these extended momenta,
one def\/ines an extended Hamiltonian,
\begin{gather}
{\cal H}(\pi_{p},\pi_{q},p,q)=\dot{p}\pi_{p}+\dot{q}\pi_{q}-{\cal
L}=H(p+\pi_{q},q)-H(p,q+\pi_{p}).\label{eq5}
\end{gather}
Using the canonical quantization rule, the following postulates
are outlined:
\begin{itemize}\itemsep=0pt
  \item  Let $p$, $q$, $\pi_{p}$ and $\pi_{q}$ be operators in a Hilbert
  space, $X$, of all complex functions, satisfying the following
  commutation relations
  \begin{gather}
  [\pi_{q},q]=-i\hbar,\qquad\pi_{q}=-i\hbar\frac{\partial}{\partial q},\qquad 
  [\pi_{p},p]=-i\hbar,\qquad\pi_{p}=-i\hbar\frac{\partial}{\partial p},\nonumber\\ 
  [p,q]=[\pi_{p},\pi_{q}]=[p,\pi_{q}]=[q,\pi_{p}]=0.\label{eq6} 
  \end{gather}
  By virtue of equations~\eqref{eq6}, the extended Hamiltonian, ${\cal H}$ will
  also be an operator in $X$.
  \item A state function $\chi(p,q,t)\in X$ is assumed to satisfy
  the following dynamical equation
  \begin{gather}
  i\hbar\frac{\partial \chi }{\partial t}={\cal
  H}\chi=\left[H\left(p-i\hbar\frac{\partial }{\partial
  q},q\right)-H\left(p,q-i\hbar\frac{\partial}{\partial p}\right)\right]\chi\nonumber\\
\phantom{i\hbar\frac{\partial \chi }{\partial t}}{}
=\sum\frac{(-i\hbar)^{n}}{n!}\left\{{\frac{\partial^{n}H}{\partial
p^{n}}\frac{\partial^{n}}{\partial
q^{n}}-\frac{\partial^{n}H}{\partial
q^{n}}\frac{\partial^{n}}{\partial p^{n}}}\right\}\chi.\label{eq7}
  \end{gather}
  \item The averaging rule for an observable $O(p,q)$, a $c$-number
  operator in this formalism, is given as
  \begin{gather*}
\langle O(p,q)\rangle=\int O(p,q)\chi^{*}(p,q,t)dpdq.
  \end{gather*}
  To f\/ind the solutions for equation~\eqref{eq7} one may assume
  \begin{gather}
  \chi(p,q,t)=F(p,q,t)e^{-\frac{ipq}{\hbar}}.\label{eq9}
  \end{gather}
  The phase factor comes out due to the total derivatives in the
  Lagrangian of equation~\eqref{eq1}, $-\frac{d(pq)}{dt}$. Substituting equation~\eqref{eq9}
  in equation~\eqref{eq7} and eliminating the exponential factor gives
  \begin{gather}
  H\left(-i\hbar\frac{\partial}{\partial q},q\right)- H\left(p,-i\hbar\frac{\partial}{\partial
  p}\right)=i\hbar\frac{\partial F}{\partial t}.\label{eq10}
  \end{gather}
  Equation \eqref{eq10} has separable solutions of the form
  \begin{gather*}
  F(p,q,t)=\psi(q,t)\phi^{*}(p,t),
  \end{gather*}
  where $\psi(q,t)$ and $\phi(p,t)$ are the solutions of the
Schr\"{o}dinger equation in the $q$ and $p$ representations,
  respectively.
\end{itemize}

\section{The extended canonical transformation}\label{sec3}

The canonical transformations that leaves the extended Hamilton
equations form invariant are the extended canonical
transformations~\cite{Sobouti}.
 Let us consider the following linear
transformation
\begin{gather}
p\rightarrow p'=p+\alpha
\frac{\hbar^{2}}{f^{2}}\pi_{p}+\beta\pi_{q},\qquad
\pi_{p}\rightarrow \pi_{p'}=\pi_{p},\nonumber\\
q\rightarrow q'=q+\alpha\pi_{q}+\beta\pi_{p},\qquad
\pi_{q}\rightarrow\pi_{q'}=\pi_{q},\label{eq12}
\end{gather}
where $\alpha$ and $\beta$ are parameters to be determined and $f$
is a positive parameter. The corresponding generator is
\begin{gather*}
G(\pi_{q},\pi_{p})=G_{1}(\pi_{q},\pi_{p})+G_{2}(\pi_{q},\pi_{p})
=\left(\frac{\hbar^{2}}{f^{2}}\frac{1}{2}\pi_{p}^{2}+\frac{1}{2}\pi_{q}^{2}\right)+(\pi_{p}\pi_{q}),
\end{gather*}
and, the corresponding similarity transformation for the f\/inite
parameters $\alpha=-\frac{f}{2i\hbar}$ and $\beta=\frac{1}{2}$
becomes~\cite{Lee}
\begin{gather}
\hat{T}=e^{i\alpha\frac{\hat{G_{1}}}{\hbar}+i\beta\frac{\hat{G_{2}}}{\hbar}}=e^{\frac{\hbar^{2}}{4f}\frac{\partial^{2}}{\partial
p^{2}}+\frac{f}{4}\frac{\partial ^{2}}{\partial
q^2}-\frac{i\hbar}{2}\frac{\partial^2}{\partial q
\partial p}}.\label{eq14}
\end{gather}
It can be easily shown that the Wigner representation can be
obtained by a canonical transformation or by the corresponding
unitary transformation in the EPS~\cite{Sobouti}. The same
technique could be applied to obtain the Wigner equation from the
Schr\"{o}dinger equation in the phase space~\cite{de Gosson}
showing the relation between the EPS technique and the Later one.
The above transformation is identical with a canonical
transformation in the Wigner representation as follows
\begin{gather*}
P\rightarrow p'=P+\alpha\frac{\hbar^{2}}{f^{2}}\pi_{P},\qquad
\pi_{P}\rightarrow \pi_{p'}=\pi_{P},\qquad Q\rightarrow
q'=Q+\alpha\pi_{Q},\qquad
\pi_{Q}\rightarrow\pi_{q'}=\pi_{Q},
\end{gather*}
where the corresponding generator is
\begin{gather*}
G'(\pi_{Q},\pi_{P})=\left(\frac{1}{2}\frac{\hbar^{2}}{f^{2}}\pi_{P}^{2}+\frac{1}{2}\pi_{Q}^{2}\right).
\end{gather*}
The corresponding similarity transformation becomes
\begin{gather*}
\hat{T'}=e^{i\alpha\frac{\hat{G'}}{\hbar}}=e^{\frac{\hbar^{2}}{4f}\frac{\partial^{2}}{\partial
P^{2}}+\frac{f}{4}\frac{\partial ^{2}}{\partial Q^2}}.
\end{gather*}
The set of $p'$, $q'$, $\pi_{p'}$, $\pi_{q'}$ now constitutes the
Husimi representation~\cite{Jannusiss}. The corresponding
operation $\hat{T'}$ on the Wigner equation will give the
evolution equation of the Husimi positive functions as follows
\begin{gather*}
\hat{T'}\left(i\hbar\frac{\partial P_{w}}{\partial
t}\right)=\hat{T'}({\cal
H}_{w}P_{w}),
\end{gather*}
where ${\cal H}_{w}$ and $P_{w}$ are the Wigner operator and the
Wigner function. Def\/ining ${\cal H}_{h}=\hat{T'}{\cal
H}_{w}\hat{T'}^{-1}$ as the Husimi operator and
$P_{h}=\hat{T'}P_{w}$ as the Husimi function~\cite{Husimi}, one
obtains the Husimi equation as follows
\begin{gather}
i\hbar\frac{\partial P_{h}}{\partial t}={\cal
H}_{h}P_{h}.\label{eq19}
\end{gather}
It can be easily shown that the Husimi and Wigner functions are
related by the following integral transform
\begin{gather}
P_{h}(p',q',t)=\frac{1}{\pi\hbar}\int dQ\int dP
e^{-\frac{(Q-q')^{2}}{f}-\frac{f(P-p')^{2}}{\hbar^{2}}}P_{w}(P,Q,t).\label{eq20}
\end{gather}
Equation \eqref{eq20}, as the Husimi function, def\/ines a class
of non-negative functions. In the case of harmonic oscillator or
radiation f\/ield, equation~\eqref{eq20} for
$f=\frac{\hbar}{m\omega}$ reduces to the well known
$Q$-function~\cite{Lee}. We will come back to this point later in
Section~\ref{sec5}.

\section{Quantum potential and generalization to the EPS}\label{sec4}

To generalize the concept of the quantum potential into EPS we
have to obtain a~$p$ space counterpart of the quantum potential.
However, unlike the~$q$ space, it does not have a simple form for
a general potential $V(q)$ in $p$ space (see details in
\cite{Brown} and~\cite{Nasiri}). As an example we consider the
harmonic potential and obtain the modif\/ied Hamilton--Jacobi
equation and quantum potential term.

The extended Hamiltonian of equation~\eqref{eq5} for the harmonic
potential, $V(q)=\frac{1}{2}kq^{2}$, becomes
\begin{gather*}
{\cal
H}=\frac{\pi_{q}^{2}}{2m}+\frac{p}{m}\pi_{q}-\frac{1}{2}k\pi_{p}^{2}-kq\pi_{p}.
\end{gather*}
The state function $\chi$ in equation~\eqref{eq7} is, in general,
a complex function. Thus we assume
\begin{gather*}
\chi(p,q,t)={\cal R}(p,q,t)e^{\frac{i{\cal
S}(p,q,t)}{\hbar}},
\end{gather*}
where ${\cal R}(p,q,t)$ is the amplitude and ${\cal S}(p,q,t)$ is
the phase given by
\begin{gather*}
{\cal S}(p,q,t)=\int^{t}{\cal
L}(p,q,\dot{p},\dot{q}, t')d t',
\end{gather*}
where ${\cal L}(p,q,\dot{p},\dot{q},t')$ is def\/ined by
equation~\eqref{eq1}. Using equation~\eqref{eq1}, one gets
\begin{gather*}
{\cal S}(p,q,t)={\cal S}^{p}+{\cal S}^{q}-pq,
\end{gather*}
where ${\cal S}^{p}$ and ${\cal S}^{q}$ are def\/ined as follows,
respectively~\cite{Nasiri}
\begin{gather*}
{\cal S}^{q}(q,t)=\int^{t}{\cal
L}^{q}(q,\dot{q},t')d t',\qquad 
{\cal S}^{p}(p,t)=\int^{t}{\cal
L}^{p}(p,\dot{p},t')d t'.
\end{gather*}
Equation \eqref{eq7} now gives
\begin{gather}
i\hbar\left(\frac{\partial {\cal R}}{\partial t}+\frac{i{\cal
R}}{\hbar}\frac{\partial S}{\partial
t}\right)=-\frac{\hbar^{2}}{2m}\left[\frac{\partial ^{2}{\cal
R}}{\partial q^{2}}+\frac{2i}{\hbar}\frac{\partial {\cal
R}}{\partial q}\frac{\partial {\cal S}}{\partial q}+\frac{i{\cal
R}}{\hbar}\frac{\partial ^{2}{\cal S}}{\partial q^{2}}-\frac{{\cal
R}}{\hbar^{2}}\left(\frac {\partial {\cal
S}}{\partial q}\right)^{2}\right]\nonumber\\
\qquad{}-\frac{i\hbar p}{m}\left(\frac{\partial {\cal R}}{\partial
q}+\frac{i{\cal R}}{\hbar}\frac{\partial {\cal S}}{\partial
q}\right)+\frac{k\hbar^{2}}{2}\left[\frac{\partial ^{2}{\cal
R}}{\partial p^{2}}+\frac{2i}{\hbar}\frac{\partial {\cal
R}}{\partial p}\frac{\partial {\cal S}}{\partial p}+\frac{i{\cal
R}}{\hbar}\frac{\partial ^{2}{\cal S}}{\partial p^{2}}-\frac{{\cal
R}}{\hbar^{2}}\left(\frac {\partial {\cal
S}}{\partial p}\right)^{2}\right]\nonumber\\
\qquad{}+i\hbar kq\left(\frac{\partial {\cal R}}{\partial
p}+\frac{i{\cal R}}{\hbar}\frac{\partial {\cal S}}{\partial
p}\right).\label{eq26}
\end{gather}
Assuming
\begin{gather*}
\pi_{p}=\frac{\partial {\cal S}(p,q,t)}{\partial p},
\qquad
\pi_{q}=\frac{\partial {\cal S}(p,q,t)}{\partial q}.
\end{gather*}
The real part of equation~\eqref{eq26} gives
\begin{gather}
\frac{\partial {\cal S}}{\partial
t}-\frac{\hbar^{2}}{2m}\frac{1}{{\cal R}}\frac{\partial ^{2}{\cal
R}}{\partial q^{2}}+\frac{\hbar^{2}k}{2}\frac{1}{{\cal
R}}\frac{\partial ^{2}{\cal R}}{\partial p^{2}}+{\cal
H}=0,\label{eq28}
\end{gather}
equation \eqref{eq28} is the modif\/ied Hamilton--Jacobi equation
for the harmonic potential in the EPS. The second and third terms
in equation~\eqref{eq28} together def\/ine the quantum potential
in extended phase space. The second term is the EPS counterpart of
quantum potential in $q$ space and the third term is the same
thing in $p$ space.

\section{Quantum potential-free representation}\label{sec5}

It was shown that in the Wigner representation the quantum
potential was removed from the corresponding modif\/ied
Hamilton--Jacobi equation for the linear and harmonic
potentials~\cite{Nasiri}. Here, we look for another possible
representation in which the quantum potential could be removed.
With the use of the canonical transformation
equation~\eqref{eq12}, its corresponding generator
equation~\eqref{eq14} and the Baker--Hausdorf\/f relation, the new
extended Hamiltonian for the Husimi function becomes
\begin{gather*}
{\cal
H}_{h}=\exp\left[-\frac{f}{4\hbar^{2}}\pi_{Q}^{2}-\frac{1}{4f}\pi_{P}^{2}\right]{\cal
H}_{w}\exp\left[-\left(-\frac{f}{4\hbar^{2}}\pi_{Q}^{2}-\frac{1}{4f}\pi_{P}^{2}\right)\right]\\
\phantom{{\cal H}_{h}}{} ={\cal H}_{w}
+i\left[-\frac{f}{4\hbar^{2}}\pi_{Q}^{2}-\frac{1}{4f}\pi_{P}^{2},{\cal
H}_{w}\right]\\
\phantom{{\cal H}_{h}=}{}
+\frac{i^{2}}{2!}\left[-\frac{f}{4\hbar^{2}}\pi_{Q}^{2}-\frac{1}{4f}\pi_{P}^{2},
\left[-\frac{f}{4\hbar^{2}}\pi_{Q}^{2}-\frac{1}{4f}\pi_{P}^{2},{\cal
H}_{w}\right]\right]+\cdots.
\end{gather*}
This Hamiltonian for the harmonic
potential, $V(q)=\frac{1}{2}kq^{2}$ becomes
\begin{gather*}
{\cal
H}_{h}=\frac{P}{m}\pi_{Q}-kQ\pi_{P}+\left(\frac{i\hbar}{2mf}-\frac{ifk}{2\hbar}\right)\pi_{Q}\pi_{P}. 
\end{gather*}
To have ${\cal H}_{h}$ in terms of the prime coordinates, one must
follow the operator transformation rule. (See the paragraph before
equation~\eqref{eq19} and the Section~5.5 in~\cite{Sobouti}.) The
above Hamiltonian for $f=\frac{\hbar}{m\omega}$ changes to the
$Q$-function Hamiltonian, ${\cal H}_{Q}$. It can be easily shown
that the modif\/ied Hamilton--Jacobi equation for this Hamiltonian
becomes
\begin{gather}
{\cal H}_{Q}+\frac{\partial{\cal S}_{Q}}{\partial
t}=0,\label{eq31}
\end{gather}
where ${\cal S}_{Q}$ is def\/ined as $Q$-action. Equation
\eqref{eq31} has the familiar form of the classical
Hamilton--Jacobi equation. As noted before, equation~\eqref{eq14}
transforms equation~\eqref{eq7} to the Husimi equation which for
$f=\frac{\hbar}{m\omega}$ the Husimi equation transforms to the
evolution equation for the $Q$-function~\cite{Lee}. Thus, we
conclude that equation~\eqref{eq31} in which the quantum potential
disappears is in the $Q$ representation.

\section{Conclusions}\label{sec6}

In this paper  the concept of  the quantum potential  in
conf\/iguration space is generalized for the extended phase space.
The Husimi  functions are obtained  by appropriate canonical
transformations in the extended phase space. By means of the EPS
formalism it is shown that the quantum potential  is removed  from
the dynamical equation of  the distribution function  in  the
Husimi  representation. Thus, the
 Hamilton--Jacobi  equation takes its standard form in the
extended phase space by excluding the quantum potential. Removing
the quantum potential in  the Husimi  representation f\/ixes the
value of parameter involved and gives the well known $Q$-function.
In addition to the present work, in a previous paper it was shown
that the quantum potential was removed for the constant, the
linear and the harmonic potentials in the Wigner
representation~\cite{Nasiri}. Now we are planning to look for the
representations where the quantum potential would be removed in
the general case.

\appendix
\section{Appendix}
To obtain the dif\/ferential form of equation~\eqref{eq20}, one
can do as follows: we change the variables as
\begin{gather*}
Q-q'=q,\qquad P-p'=p,
\end{gather*}
using the above variables, and one can rewrite
equation~\eqref{eq20} as follows:
\begin{gather}
P_{h}(p',q',t)=\frac{1}{\pi\hbar}\int dq\int dp
e^{-\frac{q^{2}}{f}-\frac{fp^{2}}{\hbar^{2}}}P_{w}(p+p',q+q',t).\notag
\end{gather}
Now using the Taylor expansion, we obtain:
\begin{gather*}
\int dq\int dp
e^{-\frac{q^{2}}{f}-\frac{fp^{2}}{\hbar^{2}}}\Bigg\{P_{w}(p',q',t)+q\frac{\partial
P_{w}(p,q,t)}{\partial q}\Big|_{q=q',p=p'}+p\frac{\partial
P_{w}(p,q,t)}{\partial
p}\Big|_{q=q',p=p'} \\
\qquad{}+\frac{q^{2}}{2}\frac{\partial^{2} P_{w}(p,q,t)}{\partial
q^{2}}\Big|_{q=q',p=p'}+ \frac{p^{2}}{2}\frac{\partial^{2}
P_{w}(p,q,t)}{\partial
p^{2}}\Big|_{q=q',p=p'}+\cdots\Bigg\}\\
\qquad{}=P_{w}(p',q',t)+ \frac{f}{4}\frac{\partial^{2}
P_{w}(p,q,t)}{\partial q^{2}}\Big|_{q=q',p=p'}+
\frac{\hbar^{2}}{4f}\frac{\partial^{2} P_{w}(p,q,t)}{\partial
p^{2}}\Big|_{q=q',p=p'}+\cdots\\
\qquad{}=e^{\frac{f}{4}\frac{\partial ^{2}}{\partial
q^{2}}+\frac{\hbar^{2}}{4f}\frac{\partial^{2}}{\partial
p^{2}}}P_{w}(p',q',t).
\end{gather*}

\subsection*{Acknowledgements}

The f\/inancial support of Research Council of Zanjan University,
Zanjan, Iran is appreciated.

\pdfbookmark[1]{References}{ref}
\LastPageEnding

\end{document}